\documentclass[showkeys,12pt,aps,prd]{revtex4-2}
\usepackage[utf8]{inputenc}
\usepackage[T1]{fontenc}
\usepackage{lmodern}
\usepackage{epsfig}
\usepackage{array}
\usepackage{amsmath}
\usepackage{epstopdf}
\usepackage{graphicx}
\graphicspath{/}

\begin{document}

\title{Zero of the proton electric form factor in the finite value of the spacelike momentum transfer squared}

\author{Erik Barto\v s}
\affiliation{Institute of Physics, Slovak Academy of Sciences,%
	Dúbravská cesta 9, 845 11 Bratislava, Slovak Republic}

\author{Stanislav Dubni\v cka}
\affiliation{Institute of Physics, Slovak Academy of Sciences,%
	Dúbravská cesta 9, 845 11 Bratislava, Slovak Republic}

\author{Noel Vincze}
\affiliation{Institute of Physics, Slovak Academy of Sciences,%
	Dúbravská cesta 9, 845 11 Bratislava, Slovak Republic}

\author{Andrej Liptaj}
\affiliation{Institute of Physics, Slovak Academy of Sciences,%
	Dúbravská cesta 9, 845 11 Bratislava, Slovak Republic}

\author{Anna Zuzana Dubni\v ckov\'a}
\affiliation{FMFI UK, Mlynská dolina F1, 842 48 Bratislava, Slovak Republic}

\date{\today}

\begin{abstract}
The behavior of the polarization data on $\mu_p(G^p_E(t)/G^p_M(t))$ ratio, at the present day available up to $t=-8.5$ GeV$^2$, obtained in the Jefferson Lab experiments carried out by the Akhiezer-Rekalo polarization method, demonstrates a decreasing tendency, which indicates the possible existence of a zero of the proton electric form factor $G^p_E(t)$ in the finite spacelike momentum transfer squared value. However, recently appeared two papers, investigating these data, one of them also included all other existing data on the nucleon electromagnetic structure in the spacelike and timelike regions, where the existence of such zero was questioned. Nevertheless, the reinvestigations we provide in this paper by means of the Unitary and Analytic model of the nucleon electromagnetic form factors give arguments for the possible existence of such a zero.
\end{abstract}

\maketitle

\section{Introduction}

While electrons in electromagnetic interactions behave as structureless objects, the protons and neutrons on the other hand, have internal structure. The latter property of nucleons is quantified by the nucleon electromagnetic (EM) structure, for the first time revealed for the proton in the elastic scattering of unpolarized electrons from unpolarized protons, $e^- p \to e^- p$ \cite{Hofst}, before the quark structure of strongly interacting particles, hadrons, has been established. This EM structure of nucleons is completely described by two independent EM form factors (FFs), the electric $G^N_E(t)$ and magnetic $G^N_M(t)$ ones, whereby the variable ``$t=-Q^2$''
\begin{equation}\label{momtrsqrt}
    t=-\frac{4E^2 \sin^2 \theta/2}{1+2E/m_N \sin^2 \theta/2}
\end{equation}
 with the incident electron energy $E$ and $\theta$ the scattering angle, is the momentum transfer squared in the process $e^- N \to e^- N$.

What experimental information on the nucleon EM FFs nowadays exists in the spacelike and timelike regions?

In the spacelike region historically the internal nucleon EM structure has been observed for the first time in the elastic scattering of unpolarized electrons on unpolarized protons by performing a straight line Rosenbluth technique \cite{Rosenbluth}, which consists of the following. In the elastic scattering process $e^- p \to e^- p$, one measures the angular distribution of the differential cross section
\begin{align}\label{difcssl}
 \frac{d\sigma^{lab}(e^- p \to e^- p)}{d\Omega}=&\frac{\alpha^2}{4E^2}\frac{\cos^2(\theta/2)}{\sin^4(\theta/2)}
 \frac{1}{1+(\frac{2E}{m_p})\sin^2(\theta/2)}\nonumber\\
&\times\Bigg[\frac{[G^p_E(t)]^2-\frac{t}{4m_p^2}[G^p_M(t)]^2}{1-\frac{t}{4m_p^2}}-2\frac{t}{4m_p^2}[G^p_M(t)]^2 \tan^2(\theta/2)\Bigg]
\end{align}
at different values of the momentum transfer squared ``$-t=Q^2$'', with $\alpha=1/137$ the fine structure constant value. The Mott differential cross-section is defined as
\begin{equation}\label{Mott}
 \frac{d\sigma}{d\Omega}_{Mott}=\frac{\alpha^2}{4E^2}\frac{\cos^2(\theta/2)}{\sin^4(\theta/2)}
 \frac{1}{1+(\frac{2E}{m_p})\sin^2(\theta/2)},
\end{equation}
then $\frac{d\sigma^{lab}(e^- p \to e^- p)}{d\Omega}/\frac{d\sigma}{d\Omega}_{Mott}$, which when evaluated for a fixed momentum transfer squared $-t$, yields a straight line when plotted against $\tan^2\theta/2$, with the slope and intercept yielding information on the both proton EM FFs.

However, using the Rosenbluth straight-line plot technique a credible information on both proton EM FFs in the spacelike region can not be extracted with increased values of $-t=Q^2$. Indeed, as one can see from (\ref{difcssl}), by increasing  $-t=Q^2$  the contribution of the proton electric $G^p_E(t)$ FF is remarkably suppressed in comparison with $G^p_M(t)$ and, as a result, the precision of the obtained information on $G^p_E(t)$ decreases.

Due to neutron instability, measurements of $G^n_E(t)$ and $G^n_M(t)$, using the Rosenbluth technique are even more difficult, hence, it is necessary to use light nuclei as ``effective'' neutron targets leading to large corrections when extracting the $e^-n \to e^-n$ cross section.

The imperfect Rosenbluth method inappropriate for $G^p_E(t)$ at higher values of the negative ``t'' is nowadays compensated by the Akhiezer-Rekalo polarization method \cite{AkhRek1,AkhRek2}, which consists in simultaneous measurements of the transverse
\begin{eqnarray}
 P_t=\frac{h}{I_0}(-2)\sqrt{\tau(1+\tau)} G^N_EG^N_M \tan(\theta/2)
\end{eqnarray}
and longitudinal
\begin{eqnarray}
 P_l=\frac{h(E+E')}{I_0m_N}\sqrt{\tau(1+\tau)} G^{N2}_M \tan^2(\theta/2)
\end{eqnarray}
components of the recoil nucleon's polarization in the electron scattering plane of the polarization transfer $\overrightarrow{e^-}N \to e^- \overrightarrow{N}$
process, with $h$ the electron beam helicity, E the beam energy, E' the final energy of electrons, $I_0$ the unpolarized cross section free from the Mott's cross section (\ref{Mott}) of the scattering of relativistic electrons on pointlike and spinless charged particle, and $\tau=Q^2/{4m^2_N}=-\frac{t}{4m^2_N}$. 

The latter method provides now a very credible information on the ratio $\frac{G^p_E(t)}{G^p_M(t)}=-\frac{P_t}{P_l}\frac{(E+E')}{2m_p}\tan(\theta/2)$,
\cite{Jones,Gayou,Punjabi,Zhan,Puckett1,Puckett2} up to $t=-8.5$ GeV$^2$.

Among other things, these data definitely disproved the validity of the universal ``dipole fit'' \cite{DChCHRWaWi} in the spacelike region
\begin{eqnarray}\label{dipole}
 G^p_E(t)\approx \frac{G^p_M(t)}{\mu_p}\approx \frac{G^n_M(t)}{\mu_n}\approx -\frac{4m^2_n}{t}\frac{G^n_E(t)}{\mu_n}\approx \frac{1}{(1-t/(0.71)^2)^2},
\end{eqnarray}
popular in the middle of 60-th of the last century, obviously derived by the zero-width vector-meson-dominance (VMD) model of a hadron ``h''
\begin{eqnarray}\label{VMD}
   F_h(s)=\sum^n_{v=1} \frac{m_v^2}{m_v^2-s}(f_{vhh}/f_v),
\end{eqnarray}
with $m_v$ the vector-meson-resonance mass, $f_{vhh}$ the coupling constant of the resonance to the considered hadron and $f_v$ the coupling constant determining the charged lepton decay width of the resonance.

In the timelike region the electric $G^p_E(s)$ and magnetic $G^p_M(s)$ proton EM FFs, where ``s'' is the total c.m. energy squared, are complex functions and experimental information on their absolute values is nowadays obtained from the experimentally measured \cite{Lees1}-\cite{Ablikim4} total cross section \cite{ZBCG}
\begin{eqnarray}\label{totcspp1}
	\sigma_{tot}(e^+e^- \to p \bar p)=\frac{4 \pi \alpha^2 C_p \beta_p(s)}{3 s}
	\Big[|G^p_M(s)|^2+\frac{2m_p^2}{s}|G^p_E(s)|^2\Big],
\end{eqnarray}
with $\beta_p(s)=\sqrt{1-\frac{4 m^2_p}{s}}$, the velocity of the outgoing proton in the c.m. system, $\alpha$=1/137, and  $C_p=\frac{\pi \alpha / \beta_p(s)}{1-\exp(-\pi \alpha / \beta_p(s))}$ in (\ref{totcspp1}) the so-called Sommerfeld--Gamov--Sakharov (SGS) Coulomb enhancement factor \cite{BPZ}, which accounts for the EM interaction between the outgoing charged $p \bar p$ pairs.

   The data not far away above the threshold $s=4m^2_p$ are practically obtained from (\ref{totcspp1}), however, by the unnatural assumption of the equality
$|G^p_E(s)|$=$|G^p_M(s)|$ exactly valid only at the threshold.

   At higher values of $s$, ignoring the insignificant value of the proton's electric FF $|G^p_E(s)|$ contribution in comparison with the proton's magnetic FF
$|G^p_M(s)|$ in (\ref{totcspp1}), the information has been extracted only on the proton's magnetic FF $|G^p_M(s)|$ with errors from the measured data on (\ref{totcspp1}).

   For a theoretical description of $|G^p_E(s)|$ and $|G^p_M(s)|$ data in the timelike region the formula (\ref{VMD}) is unusable, as the FFs behaviors there are formed
by the unstable vector meson resonance contributions, characterized not only with their masses, but also with their decay widths, which in (\ref{VMD}) is not taken into account. To solve the latter problem one needs to consider the assumed analytic properties of the proton EM FFs, which nowadays is realized by the following two approaches.

   The first one \cite{BHM} is based on an utilization of the exact analytic properties of nucleon EM FFs by means of dispersion relations of FFs with subsequent various
approximations of the spectral functions through the unitarity conditions in conformity with existing experimental data in order to achieve their most precise description.

\begin{figure}
    \includegraphics[width=0.3\textwidth]{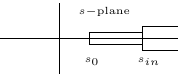}\hspace{1cm}
    \includegraphics[width=0.25\textwidth]{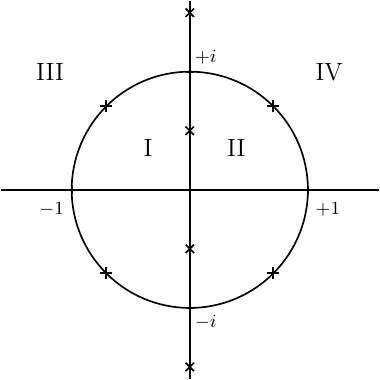}\\
\caption{Complex s-plane with square root branch points $s_0$ and $s_{in}$ (left) and conformally mapped four sheeted Riemann surface into one plane with unit disc (right).}\label{fig:1}
\end{figure}

   Our approach \cite{DD1} is based from the beginning on an utilization of an approximation of the analytic properties of the isoscalar and isovector
parts of the Dirac and Pauli nucleon EM FFs, $F_{1s}, F_{1v}, F_{2s}, F_{2v}$, to be related with $G_E(s)$ and $G_M(s)$ as follows
\begin{align}\label{relEMFFs}
G^{p,n}_E(s)=&[F_{1s}(s)\pm F_{1v}(s)]+\frac{s}{4m^2_p}[F_{2s}(s)\pm F_{2v}(s)]\\
G^{p,n}_M(s)=&[F_{1s}(s)\pm F_{1v}(s)]+[F_{2s}(s)\pm F_{2v}(s)],\nonumber
\end{align}
in the form of two square root branch points $s_0$ and $s_{in}$ on the positive real axis of the complex ``s'' plane in Fig.~\ref{fig:1} (left), whereby they are saturated by existing vector meson resonances with quantum numbers of the photon in \cite{PDG}.

The norms of these EM FFs
\begin{eqnarray}\label{normffs}
F_{1s}(0)=F_{1v}(0)=\frac{1}{2}; F_{2s}(0)=\frac{1}{2}(\mu_p+\mu_n-1); F_{2v}(0)=\frac{1}{2}(\mu_p-\mu_n-1).
\end{eqnarray}
follow from the normalization of the nucleon EM FFs
\begin{eqnarray}
G^p_E(0)=1; G^p_M(0)=\mu_p; G^n_E(0)=0; G^n_M(0)=\mu_n;
\end{eqnarray}
with $\mu_p=2.79284\mu_N$ and $\mu_n=-1.91304\mu_N$ to be the magnetic moments of the proton and neutron, respectively, expressed in the nuclear magneton units $\mu_N$
and provide for a reduction of number of free parameters in a construction of any nucleon EM structure model.

The lowest possible square root branch point ``$s_0$'' and an effective square root branch point ``$s_{in}$'' (representing contributions of all important higher inelastic thresholds into proton EM FFs) in Fig.~\ref{fig:1} (left) generate a four sheeted Riemann surface. Then the latter is by means of the function
\begin{equation}\label{confmap}
 V(s)=i\frac{\sqrt{(\frac{s_{in}-s_0}{s_0})^{1/2}+(\frac{s-s_0}{s_0})^{1/2}}-
   \sqrt{(\frac{s_{in}-s_0}{s_0})^{1/2}-(\frac{s-s_0}{s_0})^{1/2}}}
   {\sqrt{(\frac{s_{in}-s_0}{s_0})^{1/2}+(\frac{s-s_0}{s_0})^{1/2}}+
   \sqrt{(\frac{s_{in}-s_0}{s_0})^{1/2}-(\frac{s-s_0}{s_0})^{1/2}}},
\end{equation}
conformally mapped into one plane with unit disc in Fig.~\ref{fig:1} (right), where the first physical sheet of the Riemann surface is completely mapped into left half disc I., the second sheet is mapped into right half disc II., the third sheet into left half plane III. except the left half disc and finally the fourth sheet is mapped into right half plane IV. except the right half disc.

As a result the isoscalar and isovector parts of the Dirac and Pauli nucleon EM FFs, $F_{1s}, F_{1v}, F_{2s}, F_{2v}$ are expressed in the form of algebraic functions to be defined on the four sheeted Riemann surface with inclusion of contributions of complex conjugate pairs of the vector meson resonance poles, placed exclusively on unphysical sheets.

In this way the obtained Unitary and Analytic (U$\&$A) model for nucleon EM FFs represents a superposition of vector meson resonance and continuum contributions in a very natural way.

The relation (\ref{VMD}), however, does not contain any of the above mentioned properties of the U$\&$A model of the nucleon electromagnetic structure, besides the
creation of vector mesons with quantum numbers of the photon in the $e^+e^-$ annihilation into the hadron-antihadron pairs, and that also only in the primitive form considering them to be stable.

In spite of the latter, a construction of any form of the U$\&$A model of the EM structure of hadrons, including also the most accomplished U$\&$A model of the proton EM structure, is always starting from this zero-width vector-meson-dominance model (\ref{VMD}), first by an achievement of the correct asymptotic behavior, generally to be $1/s^m, 1<m\leq n$, where "n" is the number of considered resonances. With this goal, the results of the papers \cite{ADDPW} and \cite{DDW} are utilized, where first (\ref{VMD}) is transformed into a rational function with a common denominator and the polynomial of degree (n-1) in the numerator. Then requiring the first (m-1) coefficients from the highest power ``s'' in the numerator to be zero, a complicated system of linear homogeneous algebraic equations for coupling constant ratios $a_i=(f_{ihh}/f_i)$ is obtained \cite{ADDPW}.

However, utilizing the (m-1) superconvergent sum rules for the imaginary part of FFs, applying the assumed analytic properties of FFs \cite{ADDPW} and approximating $Im F(s)$ by the $\delta$-function, a more simple system of (m-1) algebraic equations is derived, which is proved \cite{ADDPW} to be equivalent to the previous one. Finally a general solution of the simpler system of (m-1) algebraic equations  \cite{DDW}, respecting normalization conditions (\ref{normffs}), for the isoscalar and isovector parts of FFs in the form of VMD model, is found. For n>m one has
\begin{align}\label{GF1}
   F_{v,s}(s)=&F_{v,s}(0) \frac{\prod^m_{j=1} m^2_j}{\prod^m_{j=1}(m^2_j-s)}+\sum^n_{k=m+1}\{\sum^m_{i=1}\frac{m^2_k}{(m^2_k-s)}.\\\nonumber
   &.\frac{\prod^m_{j\neq k,j=1}m^2_j}{\prod^m_{j\neq k,j=1}(m^2_j-s)}\frac{\prod^m_{j\neq i,j=1}(m^2_j-m^2_k)}{\prod^m_{j\neq i,j=1}(m^2_j-m^2_i)}-\\
   &-\frac{\prod^m_{j=1}m^2_j}{\prod^m_{j=1}(m^2_j-s)}\}(f_{kNN}/f_k)\nonumber
\end{align}
and for n=m the relation
\begin{eqnarray}\label{GF2}
   F_{v,s}(s)=F_{v,s}(0) \frac{\prod^n_{j=1} m^2_j}{\prod^n_{j=1}(m^2_j-s)},
\end{eqnarray}
is obtained.

In the next step all VMD terms $m^2_r/(m^2_r-s)$ in (\ref{GF1}) or (\ref{GF2}) are related to the function (\ref{confmap}) as follows.

The inverse function to (\ref{confmap})
\begin{eqnarray}\label{invrel}
    s=s_0-\frac{4(s_{in}-s_0)}{(1/V-V)^2}
\end{eqnarray}
is substituted into the VMD terms in (\ref{GF1}) or (\ref{GF2}), and by using the relations
\begin{eqnarray}
    m^2_r=s_0-\frac{4(s_{in}-s_0)}{(1/V_{r0}-V_{r0})^2}
\end{eqnarray}
and
\begin{eqnarray}
    0=s_0-\frac{4(s_{in}-s_0)}{(1/V_N-V_N)^2}
\end{eqnarray}
with $V_N=V(0)$, following from (\ref{invrel}) automatically, one gets  factorized expression

\begin{eqnarray}\label{4resp}
    \frac{m^2_r-0}{m^2_r-s}=\big(\frac{1-V^2}{1-V^2_N}\big)^2\frac{(V_N-V_{r0})(V_N+V_{r0})(V_N-1/V_{r0})(V_N+1/V_{r0})}{(V-V_{r0})(V+V_{r0})(V-1/V_{r0})(V+1/V_{r0})},
\end{eqnarray}
into the asymptotic term $\big(\frac{1-V^2}{1-V^2_N}\big)^2$ and into the so-called finite-energy term (as for $|s|\to \infty$ it turns out into the real constant), from which one deduces a creation of four independent poles. Their arrangement depends on the real part of the Breit-Wigner form squared $(m_r-i\Gamma_r/2)^2$ and if one has
\begin{eqnarray}\label{zwdth1}
   {(m^2_r+\Gamma^2_r/4)}<s_{in} \Rightarrow V_{r0}=-V^*_{r0}
\end{eqnarray}
then
\begin{eqnarray}\label{pimagax}
    \frac{m^2_r-0}{m^2_r-s}=\big(\frac{1-V^2}{1-V^2_N}\big)^2\frac{(V_N-V_{r0})(V_N-V^*_{r0})(V_N-1/V_{r0})(V_N-1/V^*_{r0})}{(V-V_{r0})(V-V^*_{r0})(V-1/V_{r0})(V-1/V^*_{r0})},
\end{eqnarray}
which produces four poles on the imaginary axis (first two inside of the unit circle and last two outside of the unit circle in Fig.1(right)). If the real part of the Breit-Wigner form squared $(m_r-i\Gamma_r/2)^2$ fulfils
\begin{eqnarray}\label{zwdth2}
   {(m^2_r+\Gamma^2_r/4)}>s_{in} \Rightarrow V_{r0}=1/V^*_{r0}
\end{eqnarray}
then
\begin{eqnarray}\label{punitcir}
    \frac{m^2_r-0}{m^2_r-s}=\big(\frac{1-V^2}{1-V^2_N}\big)^2\frac{(V_N-V_{r0})(V_N-V^*_{r0})(V_N+V_{r0})(V_N+V^*_{r0})}{(V-V_{r0})(V-V^*_{r0})(V+V_{r0})(V+V^*_{r0})}
\end{eqnarray}
which produces four poles to be symmetrically placed on the unit circle in Fig.1(right).

In the papers \cite{DD2} and \cite{ADDW}, we have analysed the actual Jefferson Lab polarization data on $\mu G^p_E(t)/G^p_M(t)$ for $-5.54$ GeV$^2 <t<-0.49$ GeV$^2$ to be considered as the most precise proton data in the spacelike region up to now, simultaneously with all electric timelike proton data, together with all spacelike and timelike magnetic proton FF data, and also electric and magnetic spacelike and timelike neutron FF data, however ignoring the questionable spacelike data on $G^p_E(t)$ for higher negative values of ``t'', obtained by the Rosenbluth technique. We used the U$\&$A model with 10 resonances, 5 isoscalars ($\omega, \phi, \omega^{\prime}, \omega^{\prime\prime}, \phi^{\prime}$) and 5 isovectors ($\rho, \rho^{\prime}, \rho^{\prime\prime}, \rho^{\prime\prime\prime}, \rho^{\prime\prime\prime\prime}$) providing 16 free parameters. A perfect description of considered data has been achieved and an existence of a zero of $G^p_E(t)$ around $t=-15$ GeV$^2$ \cite{DD2} has been found. If few experimental points with partial $\chi^2$ higher than 10 have been excluded from the analysis, the zero of $G^p_E(t)$ has been shifted to $t=-13$ GeV$^2$ in \cite{ADDW}.

In the papers \cite{DD2} and \cite{ADDW} historically the first predictions of the zero of $G^p_E(t)$ in the finite momentum transfer squared value in the spacelike region have been carried out, and they have changed the charge distribution behavior within the proton, however, leading to the value of the proton charge radius value $\sqrt<r^2_p>=0.848$ fm, which is very close to the table value $\sqrt<r^2_p>=0.841$ fm.

Later on, our prediction on the existence of the finite spacelike zero of $G^p_E(t)$ has been confirmed in the paper \cite{BBGMNP}, by describing $\mu G^p_E(t)/G^p_M(t)$ behavior exploiting proton spacelike, and timelike data and dispersion relations.

The possible existence of the zero of $G^p_E(t)$ was discussed also in the paper \cite{AdeJP}. The authors, however, view the VMD models, including also our U$\&$A modified VMD models, as requiring a significant number of free parameters to provide good fits of the data, with no credible predictive power. Thus our results on the existence of the zero of $G^p_E(t)$ from \cite{DD2} and \cite{ADDW} papers remained ignored.

Yet the predictive power of our U$\&$A models has been later on demonstrated by us, when we have analyzed the proton data as described above.

The curve we have predicted failed to follow the last three new experimental points on $\mu_p G^p_E(t)/G^p_M(t)$ between $-3.5$ GeV$^2$ and $-5.6$ GeV$^2$ (see Fig. 6 in \cite{ADDW}) and on the base of this defect our submitted paper has been refused by the redaction of the journal. But in repeated measurements in JLab the points under consideration have been corrected and the new points exactly followed our predictions. Despite the success in predicting the new fitted data points and earlier considerations by some experts, our latter results on the zero were generally unnoticed.

Recently two papers appeared \cite{TGP} and \cite{Y-HLHM}, in which the existence of the zero of $G^p_E(t)$ in the finite value of ``$-t=Q^2$'' in the spacelike region is seen questionable. We have been inspired by the latter conclusions and re-analysed the dataset in this paper using our perfected U$\&$A model with only definitely experimentally confirmed nine vector-meson resonances from \cite{PDG} and 12 free parameters, which has led us to the conclusion that there is still possibility for existence of zero in $G^p_E(t)$ in the finite spacelike region.

\section{There is still a possible zero of $G^p_E(t)$ in the finite momentum transfer squared value in the spacelike region}

In the paper \cite{TGP} the authors employ the Kuraev's model \cite{Kuraev}, based on the assumption that the center volume of the nucleon is electrically neutral due to the strong gluonic field, which has two principal effects.

The gluonic fields induce a screening which decreases the electric nucleon FF with respect to the magnetic one (beeing in conformity with the present experimental behavior of the proton polarization data on the ratio $\frac{G^p_E(t)}{G^p_M(t)}=-\frac{P_t}{P_l}\frac{(E+E')}{2m_p}\tan(\theta/2)$, \cite{Jones,Gayou,Punjabi,Zhan,Puckett1,Puckett2} up to $t=-8.5$ GeV$^2$) and moreover it favors the development of a di-quark configuration during an evaluation of the system from the quark creation to the hadron formation to be present in both spacelike and timelike regions.

On the basis of considerations discussed above, authors of the paper \cite{TGP} came to the conclusion that despite of the decreasing tendency of polarization data on the ratio $\frac{G^p_E(t)}{G^p_M(t)}=-\frac{P_t}{P_l}\frac{(E+E')}{2m_p}\tan(\theta/2)$  \cite{Jones,Gayou,Punjabi,Zhan,Puckett1,Puckett2} they can be well described by the ``monopole'' form $R=\mu_p \frac{G_{Ep}}{Mp}=(1+\frac{Q^2}{m^2_r})$ with $m^2_r=5.6 GeV^2$ not crossing zero in the finite spacelike region but vanishing or staying very small up to infinity. Whether authors of the paper \cite{TGP} are right can be revealed only by the planned extended programm of measurements at Jefferson Laboratory, following the energy upgrade \cite{Brash}, which will determine whether the ratio cross zero and eventually becomes negative.

However, the result has been obtained by the fitting of only the polarization data \cite{Jones,Gayou,Punjabi,Zhan,Puckett1,Puckett2}. It seems to us that such result on the non-existence of the zero of $G^p_E(t)$ in the finite spacelike region stands on shaky ground, as by inclusion of additional existing nucleon EM FFs experimental data in the analysis the previous result could be changed.

Results of the paper \cite{Y-HLHM} are taken into account with a pinch of salt due to the following reasons.

First, the authors carried out a simultaneous fit of data on the cross sections and ratios of proton electric to magnetic FFs in the spacelike and timelike regions, which is, in our opinion, an ill-posed problem. There is not known method up to now, how to extract from such cross sections and ratios the correct behaviors of the separate $G^p_E(t)$ and $G^p_M(t)$ FFs by a fitting. 

Second, in the paper \cite{Y-HLHM} it is written ``that 3s+3v below-threshold narrow poles and 3s+3v above-threshold broad poles were needed to obtained a good fit to the data in Table I of the paper \cite{Y-HLHM} with $\chi^2/n.d.f=0.638$''. It is not clear, how these poles are related to the experimentally confirmed vector meson resonances in \cite{PDG}.

In our opinion, as there is no theory from the first principles predicting correct behaviors of nucleon EM FFs at present, one has in a construction of any model first take into account at least all definitely experimentally confirmed vector meson resonances forming the behaviors of nucleon EM FFs and only then adjust values of few free parameters of the model by the fitting of existing data, and not to add to the model some number of isoscalar and isovector poles just as a trial to improve a fit of the data.

Further we demonstrate that if all above-mentioned features of the U$\&$A model of the nucleon EM structure \cite{DD1,ADDW} are taken into account for a description of reliable data, then the model can be adopted using only few free parameters to describe any required form given by data (not only the monopole form of the one parametric function as in the paper \cite{TGP}) and the zero of $G^p_E(t)$ in the finite momentum transfer squared value of the spacelike region can appear.

The proton and neutron EM FFs are defined by means of the isoscalar and isovector parts of the Dirac and  Pauli FFs in (\ref{relEMFFs}), which completely describe the EM structure of the proton and neutron. They can be express explicitly by means of the procedure as presented in detail above, including all 9 definitively experimentally confirmed \cite{PDG} vector mesons $\rho(770), \omega(782), \phi(1020), \rho'(1450),\omega'(1420), \phi'(1680), \rho''(1700), \omega''(1650),\phi''(2170)$ with quantum numbers of the photon $(1^{--})$. The explicit forms of $F_{1s}(s)$, $F_{1v}(s)$, $F_{2s}(s)$ and $F_{2v}(s)$ to be expressed by concrete resonances are given in Appendix A and a derivation of this form of the U$\&$A model in detail can be found in \cite{AZDD}. If the masses and widths of the considered resonances are fixed according to the PDG Table \cite{PDG}, then the model depends on 4 effective inelastic thresholds and 8 coupling constant ratios as free parameters, which are numerically evaluated (see TABLE~\ref{tab:1}) in a simultaneous fitting procedure, of the proton electric FF data \cite{Xiong} at very low momentum transfer squared ``$-t=Q^2$'' values, with the proton's MAMI separate data \cite{Bernauer}, together with new independent data on the proton $|G^p_E(s)|$ and $|G^p_M(s)|$ FFs \cite{Ablikim4}, and data on neutron $|G^n_E(s)|$ and $|G^p_M(s)|$ FFs \cite{Ablikim5}, both in the timelike region, and also together with JLab polarization data $\mu_p G^p_E(t)/G^p_M(t)$ \cite{Jones,Gayou,Punjabi,Zhan,Puckett1,Puckett2}.

An optimal description of all considered data has been obtained with the values of free parameters in Table ~\ref{tab:1}. These parameters give a rather high value of $\chi^2/ndf=2.83$, created by some of the data on the $|G^p_E(s)|$ and $|G^p_M(s)|$ FFs in the timelike region and few points of the JLab polarization data $\mu_p G^p_E(t)/G^p_M(t)$ at the highest values of ``$t=-Q^2$.

\begin{table}[tb]
	\begin{tabular}{ll}
		\hline
		$s^{1s}_{in}= (2.1451\pm 0.0057)$ GeV$^2$ & $s^{1v}_{in}= (6.1017\pm 0.0080)$ GeV$^2$\\
		$s^{2s}_{in}= (3.0532\pm 0.0168)$ GeV$^2$ & $s^{2v}_{in}= (6.2467\pm 0.0108)$ GeV$^2$\\
		$(f^{(1)}_{\omega' NN}/f_{\omega'})= -3.8618\pm 0.0249$ & $(f^{(1)}_{\phi' NN}/f_{\phi'})= 4.8090\pm 0.0289$\\
		$(f^{(1)}_{\omega NN}/f_{\omega})= -.8215\pm 0.0416$ & $(f^{(1)}_{\phi NN}/f_{\phi})= -0.0540\pm 0.0077$\\
		$(f^{(2)}_{\phi' NN}/f_{\phi'})= 1.1283\pm 0.0952$ & $(f^{(2)}_{\omega NN}/f_{\omega})= 0.9430\pm 0.1023$\\
		$(f^{(2)}_{\phi NN}/f_{\phi})= 0.0385\pm 0.0053$ & $(f^{(1)}_{\rho NN}/f_{\rho})= -0.1336\pm 0.0033$\\
		\hline
	\end{tabular}
	\caption{Values of free parameters of the proton EM structure U$\&$A model from optimal description of the proton electric EM FF data \cite{Xiong} and of the proton's  MAMI separate data \cite{Bernauer}, both measured at very low momentum transfer squared ``$-t=Q^2$'' values, together with new separate data on the proton $|G^p_E(s)|$ and $|G^p_M(s)|$ FFs \cite{Ablikim4}, and data on neutron $|G^n_E(s)|$ and $|G^p_M(s)|$ FFs \cite{Ablikim5} in the timelike region and also together with polarization data $\mu_p G^p_E(t)/G^p_M(t)$ \cite{Jones,Gayou,Punjabi,Zhan,Puckett1,Puckett2} from angular distribution measurement experiments in JLab with $\chi^2/ndf=2.83$. \label{tab:1}}
\end{table}

\begin{table}[htb]
	\begin{tabular}{ll}
		\hline
		$s^{1s}_{in}= (2.1387\pm 0.0017)$ GeV$^2$ & $s^{1v}_{in}= (5.8937\pm 0.0143)$ GeV$^2$\\
		$s^{2s}_{in}= (3.0281\pm 0.0024)$ GeV$^2$ & $s^{2v}_{in}= (5.3485\pm 0.0462)$ GeV$^2$\\
		$(f^{(1)}_{\omega' NN}/f_{\omega'})= -3.1311\pm 0.0163$ & $(f^{(1)}_{\phi' NN}/f_{\phi'})= 5.5526\pm 0.0181$\\
		$(f^{(1)}_{\omega NN}/f_{\omega})= -2.3112\pm 0.0329$ & $(f^{(1)}_{\phi NN}/f_{\phi})= 0.0059\pm 0.0195$\\
		$(f^{(2)}_{\phi' NN}/f_{\phi'})= -1.5384\pm 0.0648$ & $(f^{(2)}_{\omega NN}/f_{\omega})= 1.6733\pm 0.0580$\\
		$(f^{(2)}_{\phi NN}/f_{\phi})= -0.0050\pm 0.0086$ & $(f^{(1)}_{\rho NN}/f_{\rho})= -0.0698\pm 0.0031$\\
		\hline
	\end{tabular}
	\caption{Values of free parameters of the proton EM structure $U\&A$ model from optimal description of the proton electric EM FF data \cite{Xiong} and of the proton's  MAMI separate data \cite{Bernauer}, both measured at very low momentum transfer squared ``$-t=Q^2$'' values, without new 32 separate data on the proton $|G^p_E(s)|$ and $|G^p_M(s)|$ FFs \cite{Ablikim4}, in the timelike region, however together with the data from angular distribution measurement experiments JLab polarization data $\mu_p G^p_E(t)/G^p_M(t)$ \cite{Jones,Gayou,Punjabi,Zhan,Puckett1,Puckett2} and the neutron $|G^n_E(s)|$ and $|G^n_M(s)|$ FFs \cite{Ablikim5} in the timelike region with $\chi^2/ndf=1.63$. \label{tab:2}}
\end{table}

The polarization data are confirmed to be the most precise proton EM FF data also by the values of the partial $\chi^2$ and for such high total $\chi^2/ndf$ in TABLE~\ref{tab:1} one finds to be responsible the $|G^p_E(s)|$ and $|G^p_M(s)|$ FFs data \cite{Ablikim4} in the timelike region. Therefore we have excluded these 32 points from the analysis. The reanalysis of the remaining data with the parameter values given in TABLE~\ref{tab:2} revealed better description of them with $\chi^2/ndf=1.63$.

An explicit calculation of the behavior of $G^p_E(t)$ with parameters presented in TABLE~\ref{tab:2} has demonstrated the existence of a clear zero at $t=-22$ GeV$^2$ value.

Nevertheless, we have still found a few MAMI data points on $G^p_E(t)$ and $G^p_M(t)$ FFs \cite{Bernauer} with the partial $\chi^2$ higher value than 10, indicating a disagreement with the JLab polarization data $\mu_p G^p_E(t)/G^p_M(t)$ \cite{Jones,Gayou,Punjabi,Zhan,Puckett1,Puckett2}.

\begin{figure}[th]
\includegraphics[width=0.5\linewidth]{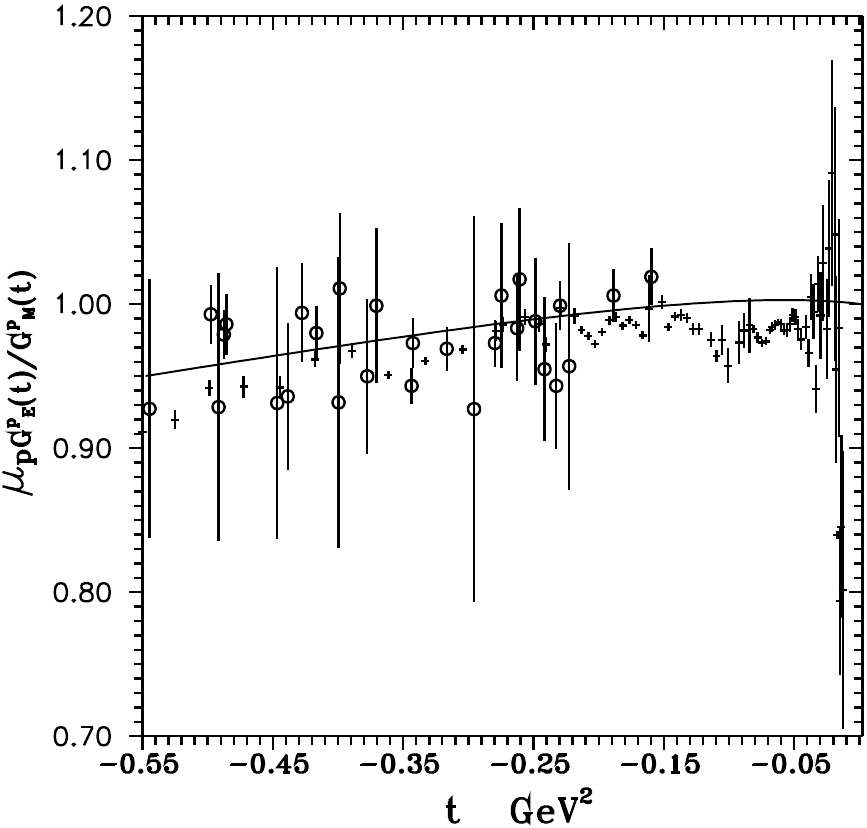}
\caption{Disagreement of few MAMI data on the proton $G^p_E(s)$ and $G^p_M(s)$ FFs, to be transformed into the form $\mu_p G^p_E(t)/G^p_M(t)$ (crosses) with JLab polarization data $\mu_p G^p_E(t)/G^p_M(t)$ data (open circles), especially for very low momentum transfer squared ``$-t=Q^2$'' values.}\label{fig:2}
\end{figure}

In order to demonstrate the latter disagreement more transparently we have transformed the MAMI data on $G^p_E(t)$ and $G^p_M(t)$ FFs separately into the ratio $\mu_p G^p_E(t)/G^p_M(t)$, with transferred errors of $G^p_E(t)$ and $G^p_M(t)$ FFs and presented them together with JLab $\mu_p G^p_E(t)/G^p_M(t)$ data
in Fig.~\ref{fig:2}.

Removing all MAMI data on $G^p_E(t)$ and $G^p_M(t)$ FFs with the partial $\chi^2$ higher value than 10 from the re-analysis one obtains the value of $\chi^2/ndf=1.227$, and the zero of $G^p_E(t)$ is shifted to the value $t=-22.4$ GeV$^2$.

The large value of the first excited state coupling constant ratios $(f^{(1)}_{\phi'NN}/f_{\phi'})$ in relation to the ground state coupling constant ratio $(f^{(1)}_{\phi NN}/f_{\phi})$ in TABLE~\ref{tab:1} and in TABLE~\ref{tab:2} is changed very little in a fitting different number of experimental data, and also the position of the zero of $G^p_E(t)$ is changed not much. So, the predicted zero seems to be stable against variations of these parameters. However, the aim of investigations in this paper is not a determination of exact position of the zero of $G^p_E(t)$, but only its existence in any finite momentum transfer squared value of the spacelike region.

There can be arisen a question. Why the obtained number of the ratio in the case of $\phi'$ is larger than in the case of the resonance $\phi$?  It could be clarified as follows. Assuming that the coupling constants $(f^{(1)}_{\phi'NN})\approx (f^{(1)}_{\phi NN})$, the lepton decay coupling constant $f_{\phi}$ is probably considerable larger than the lepton coupling constant  $f_{\phi'}$. However, there are no experimental data for the corresponding lepton decay width of $\phi$ and $\phi'$ up to now, in order to confirm or disprove our hypotheses.

\section{Conclusions}

In the papers \cite{DD2} and \cite{ADDW}, the existence of a zero around $t=-15$ GeV$^2$ \cite{DD2} and $t=-13$ GeV$^2$ \cite{ADDW} respectively, has been predicted for the first time by the analysis of current nucleon EM FF data by means of the U$\&$A models. However, recently authors of the papers \cite{TGP} and \cite{Y-HLHM} casted doubt into the existence of the zero of $G^p_E(t)$ in the finite value ``$-t=Q^2$'' of the spacelike region. We have been motivated by conclusions of these two papers. First, the arguments of such conclusion have bee re-investigated, then repeated investigations by our perfected U$\&$A model with nine definitely experimentally confirmed vector-meson resonances from \cite{PDG} were performed. These findings present clear arguments that in both cases the authors seem not to be right and the zero of $G^p_E(t)$ in the finite value ``$-t=Q^2$'' of the spacelike region has to exist. This conclusion is supported also by the result of the paper \cite{Cheng}, in which the authors independently of any model or theory of strong interaction, confirm an existence of the zero of $G^p_E(t)$ in the spacelike region even with the confidence level of $99.9\%$.

\section*{Acknowledgment}
The support of the Slovak Grant Agency for Sciences VEGA, grant No. 2/0084/25 is acknowledged.

\appendix

\section*{Appendix A}

  The most accomplished Unitary and Analytic model for the nucleon isoscalar and isovector Dirac and Pauli FFs.
\begin{eqnarray}\label{FN1s}\nonumber
  F_{1s}[V(s)]=\Bigg(\frac{1-V^2}{1-V^2_N}\Bigg)^4\Bigg\{\frac{1}{2}H_{\omega''}(V)H_{\phi''}(V)\\\nonumber
  +\Bigg[H_{\phi''}(V)H_{\omega'}(V)\frac{(C^{1s}_{\phi''}-C^{1s}_{\omega'})}{(C^{1s}_{\phi''}-C^{1s}_{\omega''})}+
  H_{\omega''}(V)H_{\omega'}(V)\frac{(C^{1s}_{\omega''}-C^{1s}_{\omega'})}{(C^{1s}_{\omega''}-C^{1s}_{\phi''})}\\
  -H_{\omega''}(V)H_{\phi''}(V)\Bigg](f^{(1)}_{\omega'NN}/f_{\omega'})\nonumber\\
  +\Bigg[H_{\phi''}(V)H_{\phi'}(V)\frac{(C^{1s}_{\phi''}-C^{1s}_{\phi'})}{(C^{1s}_{\phi''}-C^{1s}_{\omega''})}+
  H_{\omega''}(V)H_{\phi'}(V)\frac{(C^{1s}_{\omega''}-C^{1s}_{\phi'})}{(C^{1s}_{\omega''}-C^{1s}_{\phi''})}\nonumber\\
  -H_{\omega''}(V)H_{\phi''}(V)\Bigg](f^{(1)}_{\phi'NN}/f_{\phi'})\\\nonumber
  +\Bigg[H_{\phi''}(V)L_{\omega}(V)\frac{(C^{1s}_{\phi''}-C^{1s}_{\omega})}{(C^{1s}_{\phi''}-C^{1s}_{\omega''})}+
  H_{\omega''}(V)L_{\omega}(V)\frac{(C^{1s}_{\omega''}-C^{1s}_{\omega})}{(C^{1s}_{\omega''}-C^{1s}_{\phi''})}\\\nonumber
  -H_{\omega''}(V)H_{\phi''}(V)\Bigg](f^{(1)}_{\omega NN}/f_{\omega})\\\nonumber
  +\Bigg[H_{\phi''}(V)L_{\phi}(V)\frac{(C^{1s}_{\phi''}-C^{1s}_{\phi})}{(C^{1s}_{\phi''}-C^{1s}_{\omega''})}+
  H_{\omega''}(V)L_{\phi}(V)\frac{(C^{1s}_{\omega''}-C^{1s}_{\phi})}{(C^{1s}_{\omega''}-C^{1s}_{\phi''})}\\
  -H_{\omega''}(V)H_{\phi''}(V)\Bigg](f^{(1)}_{\phi NN}/f_{\phi})\Bigg\}\nonumber
\end{eqnarray}

with 5 free parameters
$(f^{(1)}_{\omega'NN}/f_{\omega'}), (f^{(1)}_{\phi'NN}/f_{\phi'}),
(f^{(1)}_{\omega NN}/f_{\omega}), (f^{(1)}_{\phi NN}/f_{\phi}),
s^{1s}_{in}$,

\begin{eqnarray}\label{FN1v}\nonumber
  F_{1v}[V(s)]=\Bigg(\frac{1-V^2}{1-V^2_N}\Bigg)^4\Bigg\{\frac{1}{2}L_\rho(V)L_{\rho'}(V)\\
  +\Bigg[L_{\rho'}(V)L_{\rho''}(V)\frac{(C^{1v}_{\rho'}-C^{1v}_{\rho''})}{(C^{1v}_{\rho'}-C^{1v}_\rho)}+
  L_\rho(V)L_{\rho''}(V)\frac{(C^{1v}_\rho-C^{1v}_{\rho''})}{(C^{1v}_\rho-C^{1v}_{\rho'})}\\\nonumber
  -L_\rho(V)L_{\rho'}(V)\Bigg](f^{(1)}_{\rho NN}/f_{\rho})\Bigg\}
\end{eqnarray}

with 2 free parameters
$(f^{(1)}_{\rho NN}/f_{\rho})$ and $s^{1v}_{in}$,

\begin{eqnarray}\label{FN2s}\nonumber
  F_{2s}[V(s)]=\Bigg(\frac{1-V^2}{1-V^2_N}\Bigg)^6\Bigg\{\frac{1}{2}(\mu_p+\mu_n-1)H_{\omega''}(V)H_{\phi''}(V)H_{\omega'}(V)\\\nonumber
  +\Bigg[H_{\phi''}(V)H_{\omega'}(V)H_{\phi'}(V)\frac{(C^{2s}_{\phi''}-C^{2s}_{\phi'})(C^{2s}_{\omega'}-C^{2s}_{\phi'})}
  {(C^{2s}_{\phi''}-C^{2s}_{\omega''})(C^{2s}_{\omega'}-C^{2s}_{\omega''})}\\\nonumber
  +H_{\omega''}(V)H_{\omega'}(V)H_{\phi'}(V)\frac{(C^{2s}_{\omega''}-C^{2s}_{\phi'})(C^{2s}_{\omega'}-C^{2s}_{\phi'})}
  {(C^{2s}_{\omega''}-C^{2s}_{\phi''})(C^{2s}_{\omega'}-C^{2s}_{\phi''})}\\\nonumber
  +H_{\omega''}(V)H_{\phi''}(V)H_{\phi'}(V)\frac{(C^{2s}_{\omega''}-C^{2s}_{\phi'})(C^{2s}_{\phi''}-C^{2s}_{\phi'})}
  {(C^{2s}_{\omega''}-C^{2s}_{\omega'})(C^{2s}_{\phi''}-C^{2s}_{\omega'})}\\\nonumber
  -H_{\omega''}(V)H_{\phi''}(V)H_{\omega'}(V)\Bigg](f^{(2)}_{\phi'NN}/f_{\phi'})\\\nonumber
  +\Bigg[H_{\phi''}(V)H_{\omega'}(V)L_{\omega}(V)\frac{(C^{2s}_{\phi''}-C^{2s}_{\omega})(C^{2s}_{\omega'}-C^{2s}_{\omega})}
  {(C^{2s}_{\phi''}-C^{2s}_{\omega''})(C^{2s}_{\omega'}-C^{2s}_{\omega''})}\\\nonumber
  +H_{\omega''}(V)H_{\omega'}(V)L_{\omega}(V)\frac{(C^{2s}_{\omega''}-C^{2s}_{\omega})(C^{2s}_{\omega'}-C^{2s}_{\omega})}
  {(C^{2s}_{\omega''}-C^{2s}_{\phi''})(C^{2s}_{\omega'}-C^{2s}_{\phi''})}+\\
  +H_{\omega''}(V)H_{\phi''}(V)L_{\omega}(V)\frac{(C^{2s}_{\omega''}-C^{2s}_{\omega})(C^{2s}_{\phi'}-C^{2s}_{\omega})}
  {(C^{2s}_{\omega''}-C^{2s}_{\omega'})(C^{2s}_{\phi''}-C^{2s}_{\omega'})}\\\nonumber
  -H_{\omega''}(V)H_{\phi''}(V)H_{\omega'}(V)\Bigg](f^{(2)}_{\omega NN}/f_{\omega})\\\nonumber
  +\Bigg[H_{\phi''}(U)H_{\omega'}(U)L_{\phi}(U)\frac{(C^{2s}_{\phi''}-C^{2s}_{\phi})(C^{2s}_{\omega'}-C^{2s}_{\phi})}
  {(C^{2s}_{\phi''}-C^{2s}_{\omega''})(C^{2s}_{\omega'}-C^{2s}_{\omega''})}\\\nonumber
  +H_{\omega''}(V)H_{\omega'}(V)L_{\phi}(V)\frac{(C^{2s}_{\omega''}-C^{2s}_{\phi})(C^{2s}_{\omega'}-C^{2s}_{\phi})}
  {(C^{2s}_{\omega''}-C^{2s}_{\phi''})(C^{2s}_{\omega'}-C^{2s}_{\phi''})}\\\nonumber
  +H_{\omega''}(V)H_{\phi''}(V)L_{\phi}(V)\frac{(C^{2s}_{\omega''}-C^{2s}_{\phi})(C^{2s}_{\phi''}-C^{2s}_{\phi})}
  {(C^{2s}_{\omega''}-C^{2s}_{\omega'})(C^{2s}_{\phi''}-C^{2s}_{\omega'})}\\\nonumber
  -H_{\omega''}(V)H_{\phi''}(V)H_{\omega'}(V)\Bigg](f^{(2)}_{\phi NN}/f_{\phi})\Bigg\}
\end{eqnarray}

with 4 free parameters
$(f^{(2)}_{\phi'NN}/f_{\phi'})$, $(f^{(2)}_{\omega NN}/f_{\omega})$,
$(f^{(2)}_{\phi NN}/f_{\phi}), s^{2s}_{in}$,

 and
\begin{eqnarray}\label{FN2v}
  F_{2v}[V(s)]=\Bigg(\frac{1-V^2}{1-V^2_N}\Bigg)^6\Bigg\{\frac{1}{2}(\mu_p-\mu_n-1)L_\rho(V)L_{\rho'}(V)H_{\rho''}(V)\Bigg\}
\end{eqnarray}
dependent on only 1 free parameter $s^{2v}_{in}$.

   The lower $L$ and the higher $H$ denotations have the following explicit forms
\begin{eqnarray}
  L_r(V)=\frac{(V_N-V_r)(V_N-V^*_r)(V_N-1/V_r)(V_N-1/V^*_r)}{(V-V_r)(V-V^*_r)(V-1/V_r)(V-1/V^*_r)},\\\label{eq19}
  C^{1s}_r=\frac{(V_N-V_r)(V_N-V^*_r)(V_N-1/V_r)(V_N-1/V^*_r)}{-(V_r-1/V_r)(V_r-1/V^*_r)}, r=\omega, \phi \nonumber
\end{eqnarray}
\begin{eqnarray}
  H_l(V)=\frac{(V_N-V_l)(V_N-V^*_l)(V_N+V_l)(V_N+V^*_l)}{(V-V_l)(V-V^*_l)(V+V_l)(V+V^*_l)},\\\label{eq20}
  C^{1s}_l=\frac{(V_N-V_l)(V_N-V^*_l)(V_N+V_l)(V_N+V^*_l)}{-(V_l-1/V_l)(V_l-1/V^*_l)}, l=
  \omega'', \phi'', \omega', \phi' \nonumber
\end{eqnarray}
\begin{eqnarray}
  L_k(V)=\frac{(V_N-V_k)(V_N-V^*_k)(V_N-1/V_k)(V_N-1/V^*_k)}{(V-V_k)(V-V^*_k)(V-1/V_k)(V-1/V^*_k)},\\ \label{eq21}
  C^{1v}_k=\frac{(V_N-V_k)(V_N-V^*_k)(V_N-1/V_k)(V_N-1/V^*_k)}{-(V_k-1/V_k)(V_k-1/V^*_k)}, k=\rho'',
  \rho', \rho \nonumber
\end{eqnarray}
\begin{eqnarray}
  L_r(V)=\frac{(V_N-V_r)(V_N-V^*_r)(V_N-1/V_r)(V_N-1/V^*_r)}{(V-V_r)(V-V^*_r)(V-1/V_r)(V-1/V^*_r)},\\ \label{eq22}
  C^{2s}_r=\frac{(V_N-V_r)(V_N-V^*_r)(V_N-1/V_r)(V_N-1/V^*_r)}{-(V_r-1/V_r)(V_r-1/V^*_r)}, r=\omega, \phi \nonumber
\end{eqnarray}
\begin{eqnarray}
  H_l(V)=\frac{(V_N-V_l)(V_N-V^*_l)(V_N+V_l)(V_N+V^*_l)}{(V-V_l)(V-V^*_l)(V+V_l)(V+V^*_l)},\\ \label{eq23}
  C^{2s}_l=\frac{(V_N-V_l)(V_N-V^*_l)(V_N+V_l)(V_N+V^*_l)}{-(V_l-1/V_l)(V_l-1/V^*_l)}, l=
  \omega'', \phi'', \omega', \phi' \nonumber
\end{eqnarray}
\begin{eqnarray}
  L_k(V)=\frac{(V_N-V_k)(V_N-V^*_k)(V_N-1/V_k)(V_N-1/V^*_k)}{(V-V_k)(V-V^*_k)(V-1/V_k)(V-1/V^*_k)},\\\label{eq24}
  C^{2v}_k=\frac{(V_N-V_k)(V_N-V^*_k)(V_N-1/V_k)(V_N-1/V^*_k)}{-(V_k-1/V_k)(V_k-1/V^*_k)}, k=\rho', \rho \nonumber
\end{eqnarray}
\begin{eqnarray}
  H_{\rho''}(V)=\frac{(V_N-V_{\rho''})(V_N-V^*_{\rho''})(V_N+V_{\rho''})(V_N+V^*_{\rho''})}
  {(V-V_{\rho''})(V-V^*_{\rho''})(V+V_{\rho''})(V+V^*_{\rho''})},\\ \label{eq25}
  C^{2v}_{\rho''}=\frac{(V_N-V_{\rho''})(V_N-V^*_{\rho''})(V_N+V_{\rho''})(V_N+V^*_{\rho''})}{-(V_{\rho''}-1/V_{\rho''})(V_{\rho''}-1/V^*_{\rho''})}.\nonumber
\end{eqnarray}

\end{document}